\documentclass[article,aps,prb]{revtex4}

\usepackage{amsmath}
\usepackage{amssymb}
\usepackage{graphicx}
\usepackage{epsfig}
\usepackage{bm}
\usepackage{units}
\usepackage{subfigure}
\usepackage{hyperref}
\usepackage{breakurl}

\newcommand{\be}{\begin{equation}}
\newcommand{\ee}{\end{equation}}
\newcommand{\bea}{\begin{eqnarray}}
\newcommand{\eea}{\end{eqnarray}}
\newcommand{\ben}{\begin{enumerate}}
\newcommand{\een}{\end{enumerate}}
\newcommand{\bit}{\begin{itemize}}
\newcommand{\eit}{\end{itemize}}

\newcommand{\la}[1]{\label{#1}}
\newcommand{\Eq}[1]{Eq.~(\ref{#1})}

\newcommand{\Fig}[1]{Fig.~\ref{#1}}

\newcommand{\vv}{\boldsymbol}					

\begin{document}

\title{The mechanical career of Councillor Orffyreus, confidence man}
 
\author{Alejandro Jenkins}\email{jenkins@hep.fsu.edu}

\affiliation{High Energy Physics, Florida State University, Tallahassee, FL 32306-4350, USA}

\affiliation{Escuela de F\'isica, Universidad de Costa Rica, 11501-2060 San Jos\'e, Costa Rica}

\date{Jan.\ 2013, last revised Mar.\ 2013; to appear in Am.\ J.\ Phys.\ {\bf 81}} 

\begin{abstract}

In the early 18th century, J.~E.~E.~Bessler, known as Orffyreus, constructed several wheels that he claimed could keep turning forever, powered only by gravity.  He never revealed the details of his invention, but he conducted demonstrations (with the machine's inner workings covered) that persuaded competent observers that he might have discovered the secret of perpetual motion.  Among Bessler's defenders were Gottfried Leibniz, Johann Bernoulli, Professor Willem 's Gravesande of Leiden University (who wrote to Isaac Newton on the subject), and Prince Karl, ruler of the German state of Hesse-Kassel.  We review Bessler's work, placing it within the context of the intellectual debates of the time about mechanical conservation laws and the (im)possibility of perpetual motion.  We also mention Bessler's long career as a confidence man, the details of which were discussed in popular 19th-century German publications, but have remained unfamiliar to authors in other languages. \\



\end{abstract}

\maketitle



\section{Introduction}
\label{sec:intro}

{\footnotesize
\begin{tabbing}
\hspace{0.41 \textwidth}

\= The perpetual motion devices whose drawings add mystery to the pages of \\
\> the more effusive encyclopedias do not work either.  Nor do the metaphysical \\
\> and theological theories that customarily declare who we are and what manner \\
\> of thing the world is. \\
\` --- Jorge Luis Borges \cite{Borges}
\end{tabbing}}

The career of Councillor Orffyreus surely ranks among the strangest episodes in the history of early modern science.  Orffyreus claimed to have solved the old riddle of perpetual motion by building a wheel that could turn continuously and do useful work (such as lifting a weight or drawing water by turning an Archimedean screw) with gravity as its only source of power.  Unlike many others who, both before and after him, claimed to have found the secret of perpetual motion, Orffyreus succeeded in convincing eminent intellectual and political figures, including the great mathematician Gottfried Leibniz and the reigning monarch of the independent German state of Hesse-Kassel, that his work was valuable.  The inventor never gave an intelligible explanation of how his machines worked, but even in the 20th century respectable authors revisited his enigmatic life and argued that he might have chanced upon a secret of nature that was lost with his death in 1745.

A brief account of Orffyreus's exploits appears in Yakov Perelman's {\it Physics for Entertainment}, a very successful work of scientific popularization first published in Russia in 1913.\cite{Perelman}  Perelman took the orthodox scientific view that perpetual motion is impossible and that Orffyreus's work was fraudulent.  On the other hand, English naval officer and amateur scholar Rupert T.~Gould included a chapter on Orffyreus in his book {\it Oddities},\cite{Oddities} first published in 1928 and re-edited in 1944, in which he meticulously documented ``anomalies'' that he believed could not be explained by modern science.\cite{Gould}  More recently, historians of science have examined this affair,\cite{Crommelin,Schaffer,Werrett} but even these treatments have left out aspects of its biographical and scientific context that might clarify what Orffyreus was doing and how his claims were received and interpreted.

\section{Perpetual motion}
\label{sec:perpetual}

In the 12th century CE, the Indian astronomer Bh\=askara II proposed attaching weights to the rim of a wheel in such a way that the wheel's own turning would shift the weights, preventing the wheel from finding an equilibrium and thus causing it to turn forever.  This idea made its way into medieval Europe via an Arabic treatise.  A similar device appears in the manuscript of French draftsman Villard de Honnecourt, from the 13th century.  The problem of perpetual motion captured the European imagination and became a staple of mechanical speculation and research.\cite{White}

The persistent failure of attempts to build a working perpetual motion machine eventually gave rise, among early modern scientists, to the idea that the laws of physics actually forbid a continued cyclic action powered by gravity alone.  In the late 16th century, the Flemish mathematician and engineer Simon Stevin correctly computed the forces acting on masses moving on an inclined plane by assuming that gravity-powered perpetual motion is impossible.\cite{Stevin}  Galileo later used similar reasoning to conclude that the speed of a mass that starts from rest and moves without friction along an inclined plane depends only on the vertical distance through which the mass has descended.\cite{Galileo}

By the mid 17th century, the active pursuit of perpetual motion devices had largely become the province of mathematically unsophisticated researchers, not directly affiliated with the universities or other learned institutions, for whom it had grown into a quest comparable to the alchemists' search for the philosopher's stone.  Figure \ref{fig:Worcester} shows an ``overbalanced wheel'' conceived by the Marquess of Worcester {\it circa} 1640,\cite{Worcester-water} in which the weights on the right always hang further from the wheel's center than those on the left, giving what he believed would be a permanent gravitational torque about the axle.  Worcester claimed to have built such a wheel and demonstrated it before King Charles I of England, but he coyly abstained from alleging that it had worked.\cite{Worcester-wheel}

\begin{figure} [t]
\begin{center}
	\includegraphics[width=0.35 \textwidth]{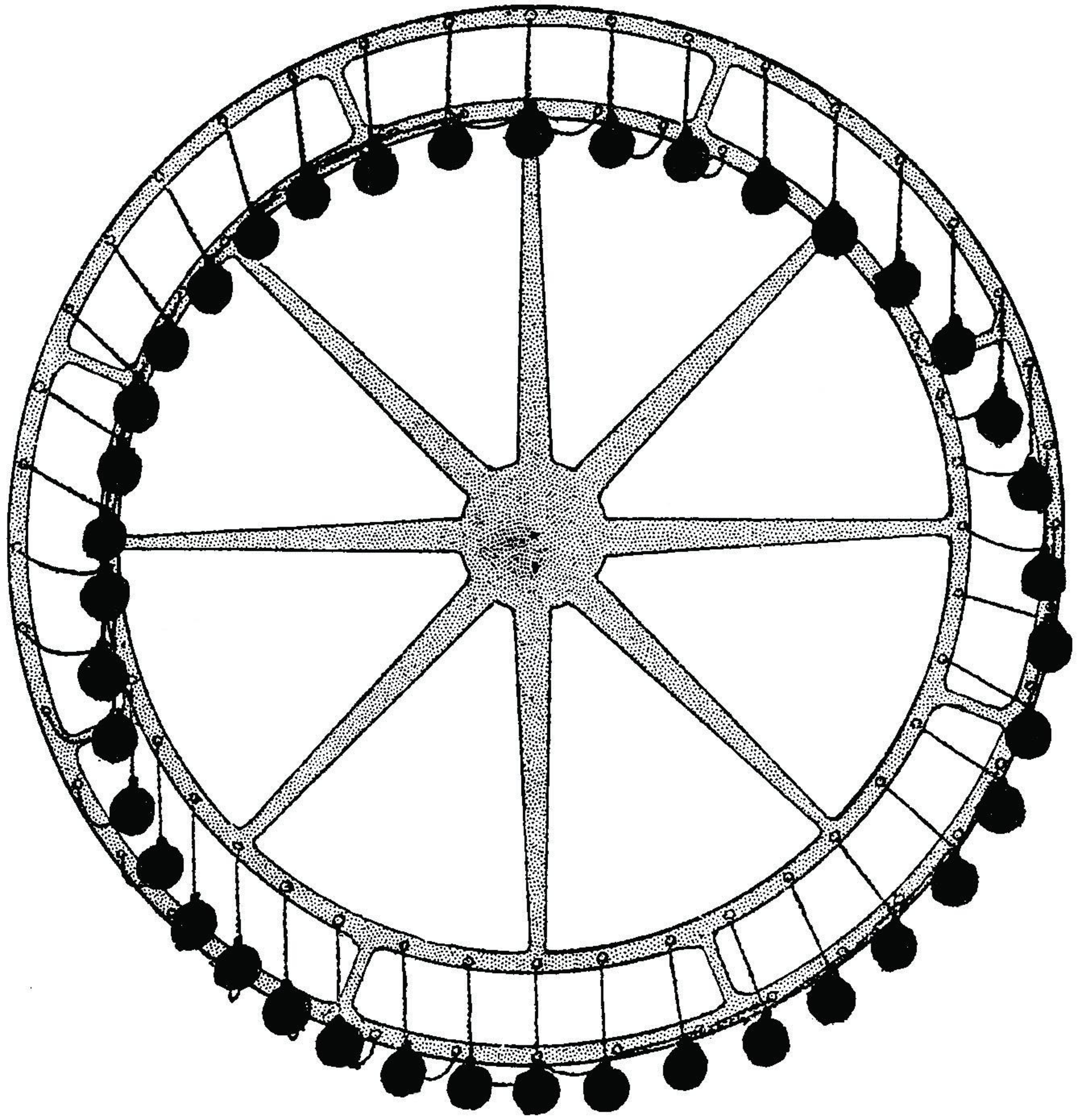}
\end{center}
\caption{Overbalanced wheel conceived by Edward Somerset, 2nd Marquess of Worcester (1601--1667).  The drawing is by Rupert T.~Gould (Fig.~14 in Ref.~\onlinecite{Oddities}) and is used here with permission of his heirs.\la{fig:Worcester}}
\end{figure}

In the 1710s, a mysterious German inventor, Councillor Orffyreus of Kassel, conducted several public demonstrations of what he claimed were overbalanced wheels that succeeded in producing perpetual motion.  In a test performed before various dignitaries and widely publicized throughout Europe, his machine was sealed in a room and then found to be turning rapidly when the room was unsealed several weeks later.  Orffyreus refused to reveal the details of his invention until he had received an adequate monetary compensation (at one point demanding \pounds 20,000, the same amount that the British Parliament had offered in 1714 for a solution to the problem of determining the longitude of a ship at sea).  In 1721 he destroyed his wheel to prevent its secret from falling into the hands of competitors.  Since then, his claims have continued to excite speculation and controversy amongst scientists and historians.

\section{\it Vis viva}
\label{sec:visviva}

In his {\it Principia Philosophi\ae} of 1644, Ren\'e Descartes emphasized the importance for the continued operation of the Universe of the law of conservation of ``quantity of motion.''  This is often presented as the first formulation of the law of conservation of momentum, but Descartes's discussion makes it clear that the quantity that he thought was conserved was not the modern vectorial
\be
\vv p = \sum_i m_i \vv v_i
\la{eq:momentum}
\ee
but rather the positive scalar
\be
q = \sum_i m_i v_i  ~,
\la{eq:Descartes}
\ee
where the $m_i$'s are the masses of particles, the $\vv v_i$'s are their velocities, and $v_i = | \vv v_i | \geq 0$ is the corresponding speed.  Descartes's reasoning was based on the belief that a vacuum is impossible and therefore that the density of matter in the Universe is constant.  Conservation of \Eq{eq:Descartes} would then follow from an equation of continuity like the one modernly derived for incompressible fluids.\cite{Descartes-visviva}

In papers presented before the Royal Society in 1668, John Wallis and Christopher Wren showed that the quantity conserved upon a collision was the $\vv p$ of \Eq{eq:momentum}, rather than the $q$ of \Eq{eq:Descartes}.  Shortly afterwards, Christiaan Huygens published a result that he had already derived in the 1650s: that elastic collisions conserve another quantity as well, corresponding to twice the modern kinetic energy,
\be
2E = \sum_i m_i v_i^2 ~.
\la{eq:Huygens}
\ee
Starting in 1686, Leibniz published several critiques of Cartesian mechanics in which he argued, following Huygens, that the conserved {\it vis viva} (``live force'') was not proportional to the speed $v$, but rather to $v^2$.  This led to a controversy with Cartesians such as the abb\'e Fran\c{c}ois de Catelan and Denis Papin.  One of the arguments that Leibniz made against the conservation of \Eq{eq:Descartes} was that it would imply the feasibility of gravity-powered perpetual motion, which (like Stevin and Galileo before him) he considered an absurdity.\cite{Leibniz-visviva}

Even after the publication of Newton's {\it Principia} in 1687, confusion persisted among Continental scientists about whether \Eq{eq:Descartes} or \Eq{eq:Huygens} gave the correct conserved quantity.\cite{vismortua}  The question was difficult to settle experimentally with the available technology and the intellectual debate was complicated by the opposition of many Newtonians to Leibniz's use of $mv^2$ since (unlike momentum $m \vv v$) it is not conserved in inelastic collisions.\cite{Leibnizians,Airy}

In the 1710s, Willem 's Gravesande, professor of mathematics and astronomy at Leiden University, argued that conservation of \Eq{eq:Descartes} meant that gravity-powered perpetual motion {\it was} possible, effectively turning Leibniz's reasoning on its head.\cite{sGravesande-perpetual}  He noted that a body that starts from rest and falls from a height $h$ achieves a speed
\be
v = \sqrt{2 g h}
\la{eq:v-height}
\ee
where $g$ is the constant gravitational acceleration.  If \Eq{eq:Descartes} were the correct expression for the conserved {\it vis viva}, then the amount of {\it vis viva} required for a mass to ascend 4 feet would be only twice the {\it vis viva} that the same mass acquires after falling from a height of 1 foot.  Thus, 's Gravesande argued that a mass could fall four times in succession, each time descending by 1 foot before hitting a solid step and transferring to it all its {\it vis viva}.  Two units of {\it vis viva} would suffice to return the mass to its original height of 4 feet, leaving another two units to overcome friction and do useful work.  This process could be repeated indefinitely, giving perpetual motion.\cite{Crommelin-perpetual}

's Gravesande, whose talents were mostly for conceiving simple experimental demonstrations of mechanical principles, then set out to estimate the {\it vis viva} of a falling mass by measuring the depth of the indentation that it leaves on a bed of wax or soft clay.  Those experiments convinced him that the {\it vis viva} was proportional to the height $h$ from which the mass falls and therefore, by \Eq{eq:v-height}, proportional to $v^2$ rather than $v$.\cite{Choc,Choc-debate}  In 1722 he therefore retracted his argument that a perpetual motion device could be powered by gravity alone.  By then he was embroiled in a public controversy about whether Councillor Orffyreus had discovered perpetual motion.\cite{Allamand,sGravesande-Hall}  

\section{Portrait of the charlatan as a young man}
\label{sec:charlatan}

The remarkable life of Councillor Orffyreus was first told in a biographical dictionary compiled in the late 18th century by Friedrich Wilhelm Strieder, the court librarian and archivist of Hesse-Kassel.\cite{Strieder}  In the 19th century, the story was repeated in popular German collections of curiosities.\cite{Roos,Buelau}  Strangely, the substance of those accounts ---which establishes that Orffyreus perpetrated a deliberate fraud--- escaped the attention of many of the authors who wrote about him in the 20th century.

Orffyreus was born Johann Ernst Elias Bessler, {\it circa} 1680, to a peasant family living in the region around the town of Zittau, in Upper Saxony.  Following a promising start as a gifted student in Zittau, the young Bessler took to the road, moving from town to town and trying his hand at many employments.  After deserting the army, he saved the life of a drowning alchemist and was rewarded with instruction on preparing elixirs.  Bessler soon established a reputation as a healer and adopted the pseudonym Orffyr\'e by writing the letters of the alphabet in a circle and choosing the letters diametrically opposite to those of his real surname.\cite{ROT13}  He then Latinized this into Orffyreus.

Orffyreus accompanied a rich gentleman to Italy, where he visited a monastery whose kitchen boasted a roasting spit that turned automatically.  He grew obsessed with the problem of perpetual motion, but poverty forced him to leave Italy and find employment as a watchmaker.  He continued his itinerant life and at one point was asked by Dr.\ Christian Schuhmann, the mayor and physician of Annaberg, to heal his desperately ill daughter.  The girl recovered and became Orffyreus's wife, greatly improving his social and financial standing.\cite{Rychlak}

\section{Orffyreus's machines}
\label{sec:wheels}

In 1712 Orffyreus reached the city of Gera, where he exhibited a small wheel that he claimed turned on its own due to an ingenious mechanism hidden within.  He built progressively larger wheels which he exhibited in Draschwitz (a small village between Zeitz and Leipzig) and later in the town of Merseburg.  Those demonstrations attracted considerable attention, including from Leibniz, who in 1714 visited Orffyreus in Draschwitz\cite{Draschwitz} and later wrote in a letter to Robert Erskine, the personal physician and scientific advisor to Russian Tsar Peter the Great, that he believed Orffyreus's wheel to be a valuable invention.\cite{Erskine}  In that letter Leibniz described Orffyreus as ``one of my friends'' and noted that he had advised testing the wheel for several weeks under rigorous surveillance in order to produce a compelling testimonial that would encourage a potentate to pay a substantial reward in exchange for its secret.\cite{Leibniz-Peter}  Orffyreus's drawing of the Merseburg wheel is shown in \Fig{fig:Merseburg}.

\begin{figure} [t]
\begin{center}
	\includegraphics[width=0.65 \textwidth]{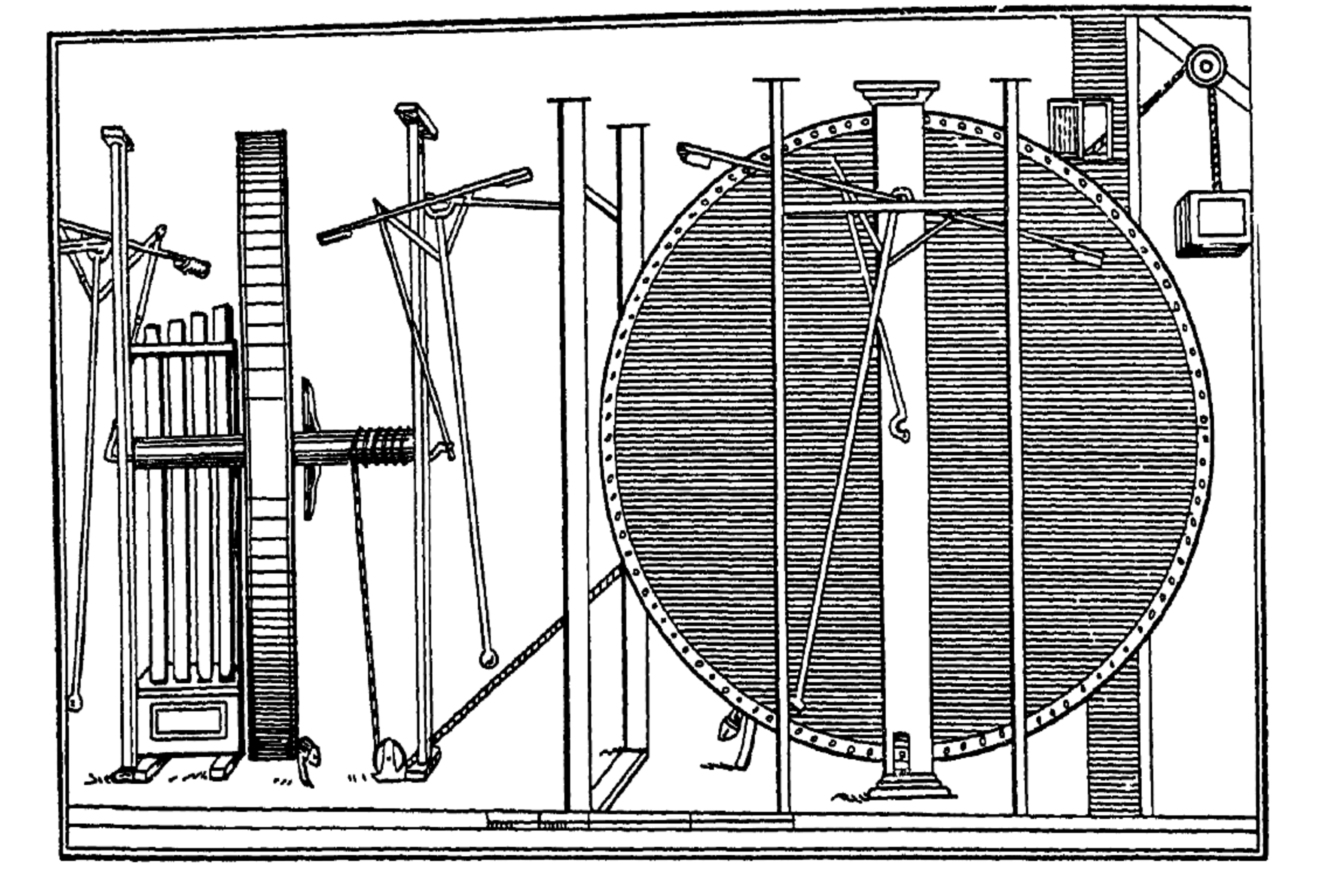}
\end{center}
\caption{Diagram of the wheel built by Orffyreus in Merseburg, taken from Ref.~\onlinecite{Perelman} and based on an engraving published by Orffyreus in 1715.  On the left is a side view and on the right a frontal view, both of the same device.  The driving mechanism is supposedly hidden within the wheel.  The three-bobbed pendula on the front and back of the wheel merely regulate its speed of rotation.  The device is shown lifting a weight attached to a rope that wraps around the axle.\la{fig:Merseburg}}
\end{figure}

Other witnesses were skeptical of Orffyreus's invention and even published reports claiming that the wheel was turned by a concealed mechanism.  Mining engineer Johann Gottfried Borlach circulated an engraving, shown in \Fig{fig:Borlach}, illustrating how he thought the machine was moved by human force from an adjacent room.\cite{Borlach}  Andreas G\"artner, the Dresden-based model-maker to the King of Poland, and Leipzig mathematician Christian Wagner also publicly denounced Orffyreus as a fraud.\cite{Dircks}  Nonetheless, Orffyreus gained the patronage of Prince Karl, the Landgrave of the independent state of Hesse-Kassel in central Germany and an enthusiast of the new sciences who maintained an extensive collection of clocks, steam engines, and other machines.  Orffyreus moved to Kassel in 1716.  There, he was appointed Commercial Councillor ({\it Kommerzialrat}) and given rooms in the ducal castle of Weissenstein, as well as a estate in nearby Karlshafen.

\begin{figure} [t]
\begin{center}
	\subfigure[]{\includegraphics[width=0.3 \textwidth]{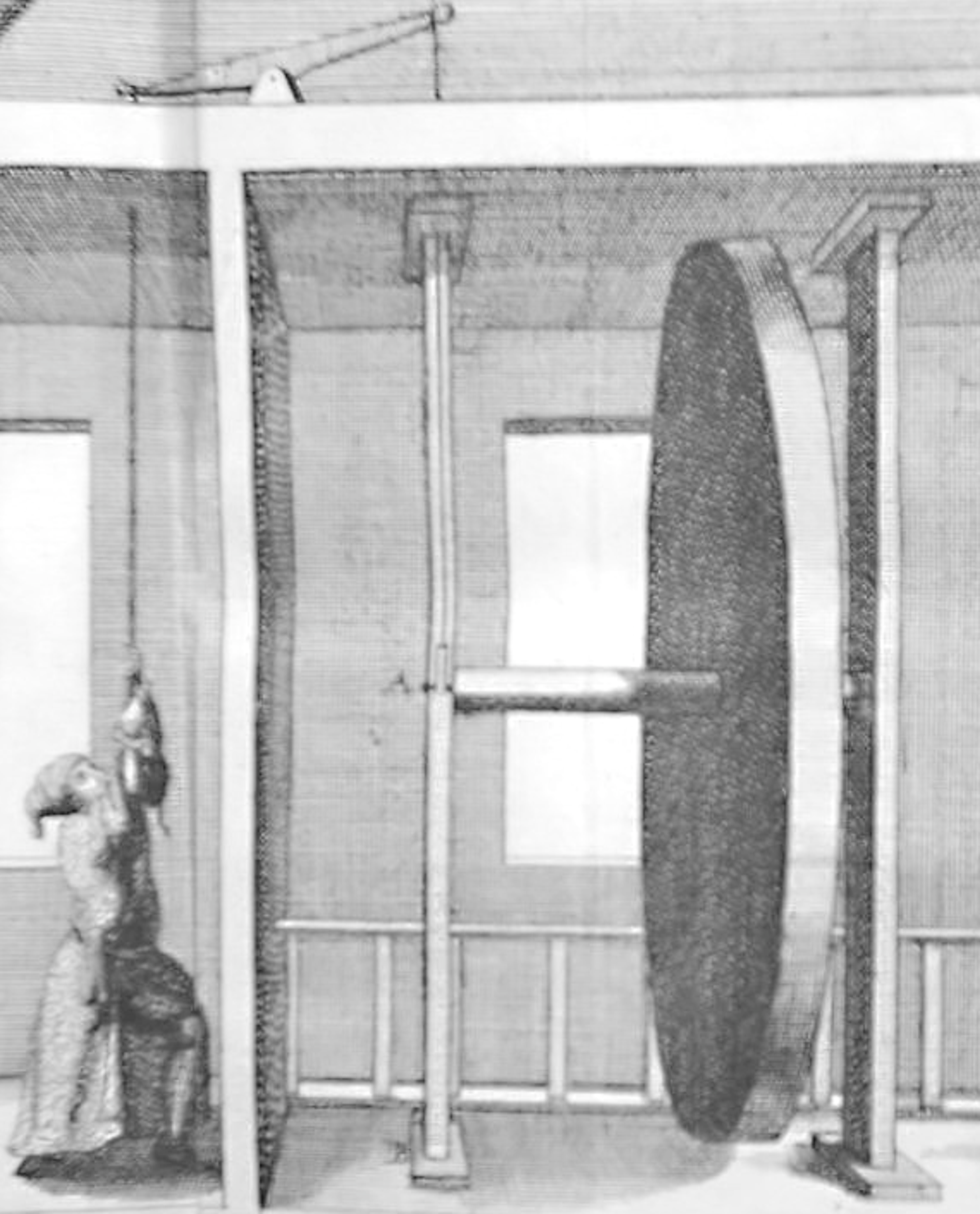}}  \hskip 2.5 cm
	\subfigure[]{\includegraphics[width=0.33 \textwidth]{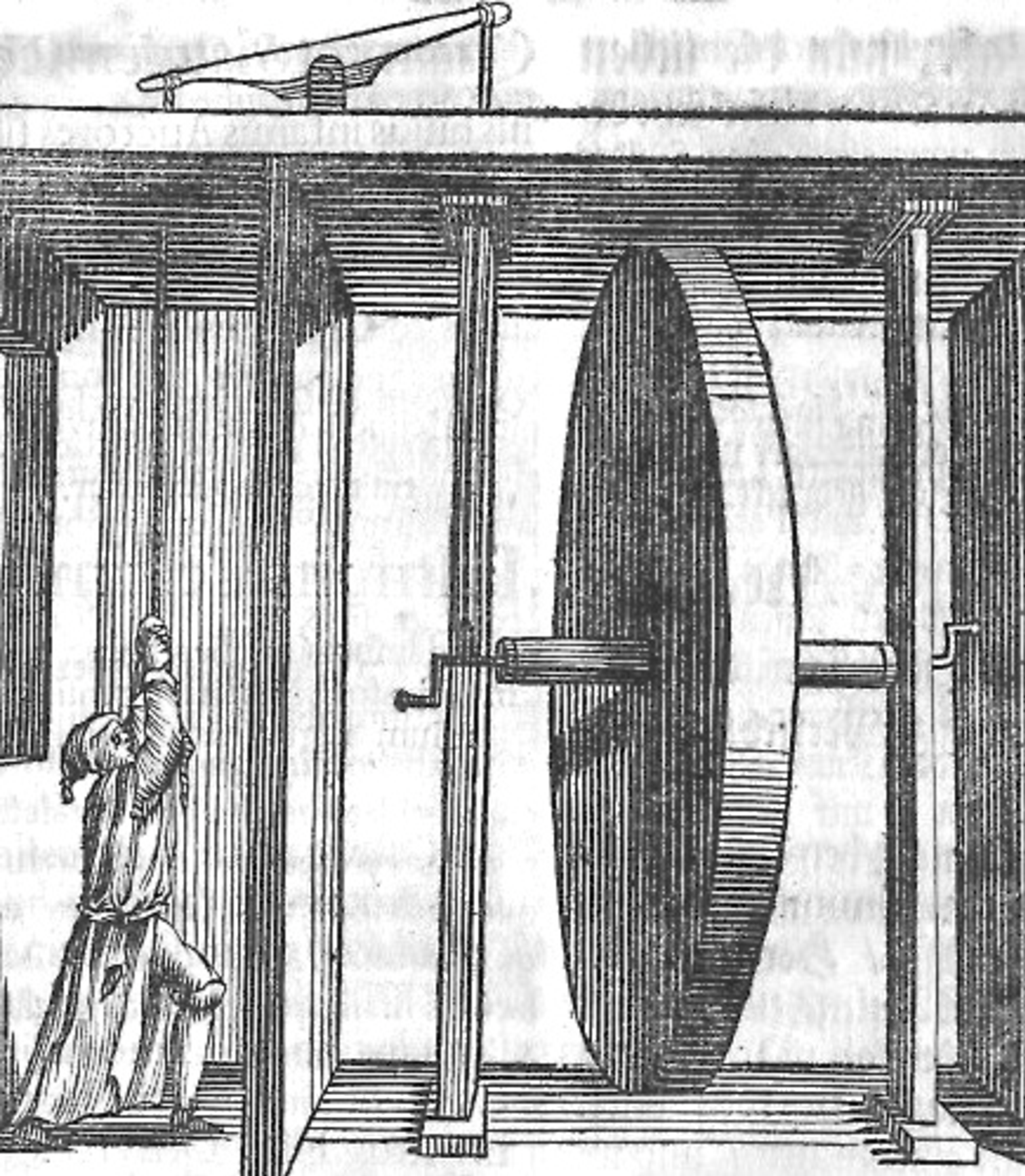}}
\end{center}
\caption{(a) Detail of an accusatory image published by engineer Johann Borlach after witnessing Orffyreus's exhibition at Merseburg.\cite{Borlach} (b) Version given by Orffyreus in Ref.~\onlinecite{Triumphans}, accompanied by a brief text dismissing Borlach's interpretation as absurd.  Suspiciously, Orffyreus altered the image in significant ways, adding handles to the axle and obscuring the crank that Borlach thought coupled the wheel to the lever.\la{fig:Borlach}}
\end{figure}

The wheel that Orffyreus built in Weissenstein was 12 feet (3.6 m) in diameter and 14 inches (36 cm) thick.  It was covered with waxed linen to conceal the internal mechanism, which only Prince Karl was ever allowed to see.  According to an account by architect Joseph Emmanuel Fischer von Erlach ``at every turn of the wheel can be heard about eight weights, which fall gently on the side towards which the wheel turns.''\cite{Fischer}  Orffyreus published extravagant pamphlets extolling his own work and offering to reveal its secrets in exchange for a monetary reward, while explaining only that the weights were cunningly arranged so that the wheel could never reach an equilibrium.\cite{Triumphans}

In 1717--18, the Landgrave conducted a public test ---much like the one suggested by Leibniz--- in which the room containing the wheel was carefully sealed for several weeks.  When the seals were broken the device was found to be turning rapidly.  's Gravesande visited Kassel in 1721 and wrote a letter to Isaac Newton expressing his conviction that the wheel was not turned by any external power.\cite{sGravesande-Newton}  Orffyreus, however, smashed his machine after another examination of the axle's exterior caused him to suspect that 's Gravesande was trying to discover the machine's secret without paying for it.  Nonetheless, 's Gravesande continued to defend Orffyreus's work even after his own experiments forced him to abandon his original argument for perpetual motion.

Meanwhile, John Desaguliers, the Royal Society's curator of experiments and a close associate of Newton, responded to the reports of Orffyreus's wheel with what effectively amounts to the modern argument that an overbalanced wheel (like the one shown in \Fig{fig:Worcester}) cannot work because a device moved only by gravity must keep lowering its center of mass in order to accelerate or to maintain its turning against friction.\cite{Desaguliers}  This follows from Galileo's principle that the maximum speed that a weight can gain from the action of gravity depends only on its net vertical descent.  Whatever kinetic energy a mass gains as it falls will be entirely lost when it returns to its original height upon completing a revolution, regardless of the path taken.\cite{Stewart}  Desaguliers conducted experimental demonstrations before the fellows of the Royal Society to support that conclusion.\cite{RS-demos}

\section{The end of the affair}
\label{sec:end}

The record of how and when Orffyreus's hoax was exposed is hazy, probably because of the natural disinclination on the part of the Hessian authorities and of the savants who had vouched for the machines to acknowledge how far they had been duped.  In a letter dated 18 May 1727, the Landgrave's Superintendent of Works informed 's Gravesande that Orffyreus was building a new wheel that would soon be ready for testing, but no news of the machine's progress followed.

A few months later, Orffyreus's maid, Anne Rosine Mauersbergerin, ran away from her employer's home and revealed that the wheels had always been turned from an adjacent room, sometimes by her, sometimes by Orffyreus's wife or by his brother Gottfried, sometimes by Orffyreus himself.  She produced a written oath that Orffyreus had threatened her into signing, in which she had sworn under pain of damnation not to reveal what she knew of the machine's operation.  It also emerged that Orffyreus's wife had already informed a government official of her husband's deception, but had been angrily instructed to keep quiet.

It is not possible now to determine with certainty how Orffyreus's wheels turned.  Mauersbergerin's testimony, as quoted in Ref.~\onlinecite{Strieder}, is unclear on the mechanical details, but it agrees with Borlach's earlier allegation that one of the posts was hollow and contained a coupling of the axle to a human force acting from an adjacent room (see \Fig{fig:Borlach}(a)).  's Gravesande dismissed the maid's claims on the grounds that they were inconsistent with what he had seen of the wheel and that he paid ``little attention to what a servant can say about machines.''\cite{Crousaz}  's Gravesande and others reported seeing the ends of the axle turning freely in the exposed bearings, but this does not really rule out the presence of a turning mechanism hidden inside the post.  With enough cleverness, Orffyreus could even have made the mechanism so that the axle could be dismounted and exhibited without giving away his secret.\cite{axle}

During Orffyreus's time in Kassel, he also tried unsuccessfully to establish an Orffyrean church, which he hoped would transcend the differences between Catholics and Protestants.  Prince Karl died in 1730 and Orffyreus was questioned and arrested in 1733, though by 1738 he was again free and living in his estate in Karlshafen, where he advertised outlandish new inventions and schemes.  In 1745 he fell to his death while constructing a windmill in F\"urstenberg.

\section{Leibniz and mixed perpetual motion}
\label{sec:mixed}

Why did Leibniz, who in his controversies with the Cartesians had dismissed gravity-powered perpetual motion as an absurdity, advertise Orffyreus's wheel in a letter to the Tsar's physician? It might be significant that that letter was written in 1716, during the last months of Leibniz's life, at a time when the priority dispute with Newton over the invention of calculus had negatively impacted his international reputation and when he had fallen out of favor with his employer, Prince Georg Ludwig of Hanover, who had acceded to the British throne in 1714 and moved to a London in which Newton dominated the scientific community as president of the Royal Society.\cite{Leibniz}  Leibniz might therefore have been eager to gain favor with the Tsar and other princes by calling attention to a potentially revolutionary German invention.

Furthermore, Johann Bernoulli, a close associate of Leibniz and an ally in his controversies with Newton, had long claimed that some sort of perpetual motion had to be possible, because otherwise the Universe would not keep running on its own.\cite{osmosis}  Bernoulli argued that microscopic ``natural'' processes (such as fermentation) could be used to power a macroscopic motion and replenish mechanical losses, as demonstrated by the continual motion of wind, water, and living creatures.\cite{Gabbey}  In 1722, Bernoulli wrote to 's Gravesande that Orffyreus's wheel was probably an implementation of a ``mixed'' perpetual motion, in which the action of gravity (which could not, on its own, maintain a cyclic motion) was complemented by some other ``active principle.''\cite{Bernoulli}

This argument had religious and political undertones because of Newton's claim, which Bernoulli and Leibniz derided, that divine intervention is necessary to maintain the planets' regular motion around the sun.\cite{Newton}  In 1715 Leibniz had written to the Princess of Wales that, according to the Newtonians,
\begin{quote}
God Almighty wants to wind up his watch from time to time: otherwise it would cease to move.  He had not, it seems, sufficient foresight to make it a perpetual motion.  Nay, the machine of God's making, is so imperfect, according to these gentlemen; that he is obliged to clean it now and then by an extraordinary concourse, and even to mend it, as a clockmaker mends his work; who must consequently be so much the more unskilful a workman, as he is oftener obliged to mend his work and set it right.\cite{Clarke}
\end{quote}
This led to an exchange of polemical letters with Samuel Clarke, a Newtonian clergyman, which was ongoing at the time Leibniz's death in 1716.  Historians of science have commented extensively on the ideological aspects of the Leibniz-Clarke correspondence,\cite{Leibnizians,Shapin} but it has not always been sufficiently appreciated that Newton's argument for continuing divine intervention did not really {\it derive} from metaphysics.  Newton had actually posed the problem of the dynamic stability of the solar system, which has continued to occupy physicists and mathematicians to this day.\cite{Tremaine}  Leibniz, who rejected the Newtonian theory of gravity, was the first to frame the issue as principally theological, in a letter intended to cause trouble for Newton with the new Hanoverian court in London.\cite{Brown}

In the context of the increasingly bitter rivalry between Leibnizians and Newtonians, it might be understandable that the former would seize on Orffyreus's claim of perpetual motion with imperfectly critical enthusiasm.  Two other of Orffyreus's early defenders were Viennese architect J.~E.~Fischer von Erlach, a pupil of Leibniz who became a pioneer in the development of steam engines, and the Liebnizian philosopher Christian Wolff, later chancellor of the University of Halle.  The Leipzig {\it Acta Eruditorum}, a learned journal co-founded by Leibniz, published three positive accounts of Orffyreus's work in 1715--18\cite{Acta} and Wolff mentioned Orffyreus in his 1716 mathematical encyclopedia.\cite{Wolff1}

After 1722, when 's Gravesande abandoned his original argument for perpetual motion based on conservation of $mv$, he converted to Bernoulli's views and contended that Orffyreus had probably mastered a new force of nature that allowed his wheels to turn continuously, an opinion that 's Gravesande maintained for the rest of his life.\cite{Crommelin}  For his part, Christian Wolff dropped all mention of Orffyreus in the later editions of his encyclopedia, but he continued to insist that mixed perpetual motion might be possible, while counseling mathematicians and natural philosophers to leave its pursuit to the practical mechanics.\cite{Wolff2}  Shortly after his death in 1754, Wolff's old article from the {\it Acta} was reprinted with a discreet parenthetical note after Orffyreus's name: {\it cujus occultationis fraudes adbuc ignoti erant} (``whose secret fraud was still unknown'').\cite{fraudes}

\section{Conclusions}
\label{sec:conclusions}

Though Orffyreus's machines met with vocal skepticism from the start, the positive response of Leibniz and his associates must have helped the inventor to gain and maintain the support of the Landgrave of Hesse-Kassel and to attract overtures from Peter the Great.  The role of the Leibnizians in the Orffyreus affair also illustrates the complicated relation between the content of scientific debates and the personal allegiances of the scientists involved in them.  

's Gravesande began as a Newtonian who believed that gravity-powered perpetual motion was possible, based on a misunderstanding of Newton's mechanics.  This led him to a conclusion that Leibniz had previously deemed absurd in his controversies with the Cartesians.  When 's Gravesande's own experiments disabused him of that idea, he joined the Leibnizians both in advocating $mv^2$ as the conserved {\it vis viva} and in believing that Orffyreus's machines used some new force of nature, despite Orffyreus's claims that they were moved only by gravity.

The Leibnizian enthusiasm for Orffyreus's machines was probably motivated in part by Leibniz's pursuit of patronage at a time when his personal fortunes had fallen considerably.  It was then sustained by the role of the question of ``mixed'' perpetual motion in the public controversies over the metaphysical implications of Newton's theory of gravity (a theory that the Leibnizians rejected).

It is remarkable that Orffyreus should have fooled such august persons for so long simply by concealing a mechanism that allowed him and his accomplices to turn the wheel from an adjacent room.  In our own time, it has been remarked that scientists are not normally trained to look for deliberate deceptions and can therefore be duped by cunning charlatans as easily as other people.\cite{Randi}  Furthermore, it is understandable that, after Orffyreus's imposture was revealed, his former supporters should have been loath to acknowledge or publicize their mistake.  This allowed even respectable authors, beginning with Jean Allamand,\cite{Allamand} to entertain the thought that Orffyreus might have taken the secret of perpetual motion to his grave, a view that Gould\cite{Oddities} and Crommelin\cite{Crommelin} repeated in the 20th century.

Bessler's early life suggests ready charm and a burning ambition to rise from his humble origins, coupled to genuine skills but also to grave flaws in his personal character and intellectual training.  It is possible that his pursuit of perpetual motion began in good faith and that he first turned to deceit in an attempt to gain the time and money that he thought he needed to perfect the real thing.  The court tutor at Kassel, Jean-Pierre de Crousaz, described Bessler as a ``madman''\cite{Crousaz} and his distraught wife spoke of him as a ``desperate man who cared about nothing,''\cite{Strieder} but his erratic and irascible behavior might be explained by the need to drive away inconvenient examinations of his work, as well as by the psychological strain of sustaining a career that had come to depend on a lie.   

Bessler bequeathed his mechanical secrets to his son-in-law, locksmith Johann Adam Crone.  In Friedrich B\"ulau's words, this ``would have been worth something only in the hands of a quacksalver.''\cite{Buelau}  Bessler's machines did excite significant interest and controversy among some of the leading scientists of the day (including 's Gravesande, Desaguliers, Leibniz, and Bernoulli).  On account of their role in the early debates about what would become the law of energy conservation, Bessler's deceptions might therefore have some place, after all, in the historical development of mechanics.

\begin{acknowledgements}

I thank Tristan McLoughlin for discussions and for pointing me to Ref.~\onlinecite{Leibniz-visviva}, and Tom Hayes for assistance in procuring a copy of Ref.~\onlinecite{Crommelin} and for suggestions on improving this manuscript.  Carola Berger prepared English translations of Refs.~\onlinecite{Roos} and \onlinecite{Buelau}, through her CFB Scientific Translations service.  Those translations are available from me upon request.  I gratefully acknowledge Carola's assistance in clarifying points of language and history and her discovery of Ref.~\onlinecite{Rychlak}.  I also thank Alex Boxer and Guido Festuccia for assistance with the Latin of Refs.~\onlinecite{Acta} and \onlinecite{fraudes}, Sarah Stacey for permission to use \Fig{fig:Worcester}, Jonathan Betts for correspondence on the career of Rupert Gould, and David Kaiser for encouragement with this project.

\end{acknowledgements}


\bibliographystyle{aipprocl}   

\begin{thebibliography}{99}
\bibitem{Crommelin-perpetual}
's Gravesande's argument has been modernly quoted and explained in Ref.~\onlinecite{Crommelin}.
\bibitem{Choc-debate}
Similar experiments had already been conducted by Giovanni Poleni, of the University of Padua, which 's Gravesande learned about only after he had carried out his own tests.  The interpretation of Poleni's and 's Gravesande's results provoked a debate between Leibnizians and Newtonians, which from our present perspective could have been resolved by noting that a constant force $\vv F$ changes the momentum of a body by an amount $\Delta \vv p = \vv F \, \Delta t$ and its energy by $\Delta E = \vv F \cdot \Delta \vv r$, where $\Delta t$ is the time interval and $\Delta \vv r$ is the displacement over which the force acts.  See Ref.~\onlinecite{Leibnizians}.
\bibitem{ROT13}
Today this procedure is called a ROT13 cipher.
\bibitem{Leibniz-Peter}
Leibniz's involvement goes unmentioned in most scholarly accounts of the Orffyreus affair.  On Peter the Great's interest in Orffyreus's wheel, the surrounding negotiations, and the need for competent scientific advice that led to the founding of the St.\ Petersburg Academy of Sciences, see Ref.~\onlinecite{Werrett}.
\bibitem{axle}
's Gravesande recalled in Ref.~\onlinecite{Crousaz} that it had been his examination of the Weissenstein machine's bearings that threw Orffyreus into such a rage that he smashed the device, but this did not arouse in the professor any suspicion that Orffyreus might have feared too close a scrutiny of the axle and its supports.  The cumbersome three-bobbed pendulums shown in \Fig{fig:Merseburg}, ostensibly needed to regulate the turning speed, might have helped divert attention during examinations of the axle.
\end{thebibliography}

\end{document}